\documentclass[useAMS,usenatbib]{mn2e}

\usepackage{graphicx}
\usepackage{aas_macros}
\usepackage{amssymb}
\usepackage[breaklinks,colorlinks=true,citecolor=black,linkcolor=black,urlcolor=blue,bookmarks=true]{hyperref}

\newcommand{\feh}{\mbox{[Fe/H]}}
\newcommand{\kms}{km\,s$^{-1}$}
\newcommand{\hi}{\textrm{H}\,\textsc{i}}
\newcommand{\hii}{\textrm{H}\,\textsc{ii}}
\newcommand{\msol}{\textrm{M}$_{\odot}$}

\newcommand{\ii}{\textsc{ii}}

\newcommand{\OIII}{[\textrm{O}\,\textsc{iii}]}
\newcommand{\HeI}{\textrm{He}\,\textsc{i}}
\newcommand{\HeII}{\textrm{He}\,\textsc{ii}}
\newcommand{\SII}{\textrm{[S}\,\textsc{ii}]}
\newcommand{\Halpha}{\textrm{H}$\alpha$}

\newcommand{\HSTphot}{\textsc{HSTphot}}

\title[A unique dSph galaxy]{A unique isolated dwarf spheroidal galaxy at D=1.9 Mpc}

\author[D.\ Makarov et al.]{
Dmitry Makarov$^{1}$\thanks{E-mail: dim@sao.ru},
Lidia Makarova$^{1}$,
Margarita Sharina$^{1}$,
Roman Uklein$^1$,
\newauthor
\framebox{Anton Tikhonov}$^{\,2}$,
Puragra Guhathakurta$^3$,
Evan Kirby$^4$
and
Natalya Terekhova$^5$
\\
$^{1}$Special Astrophysical Observatory, Nizhniy Arkhyz, Karachai-Cherkessia 369167, Russia\\
$^{2}$Saint-Petersburg State University, Russia\\
$^{3}$UCO/Lick Observatory, University of California-Santa Cruz, Santa Cruz, CA 95064, USA\\
$^{4}$California Institute of Technology, 1200 E. California Blvd., MC 249-17, Pasadena, CA 91125, USA\\
$^{5}$Sternberg Astronomical Institute, Moscow State University, Russia
}

\begin{document}

\voffset=-0.8in

\date{}
\pagerange{\pageref{firstpage}--\pageref{lastpage}} \pubyear{XXX}
\maketitle

\label{firstpage}

\begin{abstract}
We present a photometric and spectroscopic study of the unique isolated nearby dSph galaxy KKR25.
The galaxy was resolved into stars with HST/WFPC2 including old red giant branch and red clump. 
We have constructed a model of the resolved stellar populations and measured the star formation rate and metallicity as function of time. 
The main star formation activity period occurred about 12.6 to 13.7 Gyr ago. 
These stars are mostly metal-poor, with a mean metallicity $\feh \sim -1$ to $-1.6$ dex. 
About 60 per cent of the total stellar mass was formed during this event. 
There are indications of intermediate age star formation in KKR25 between 1 and 4 Gyr with no significant signs of metal enrichment for these stars. 
Long-slit spectroscopy was carried out using the Russian 6-m telescope of the integrated starlight and bright individual objects in the galaxy.
We have discovered a planetary nebula (PN) in KKR25. 
This is the first known PN in a dwarf spheroidal galaxy outside the Local Group. 
We have measured its oxygen abundance $12+\log(\textrm{O/H}) = 7.60 \pm 0.07$ dex and a radial velocity $V_h = -79$ \kms{}. 
We have analysed the stellar density distribution in the galaxy body. 
The galaxy has an exponential surface brightness profile with a central light depression. 
We discuss the evolutionary status of KKR25, which belongs to a rare class of very isolated dwarf galaxies with spheroidal morphology. 
\end{abstract}


\section{Introduction}

\begin{figure*}
\begin{tabular}{cc}
\includegraphics[height=0.45\textwidth]{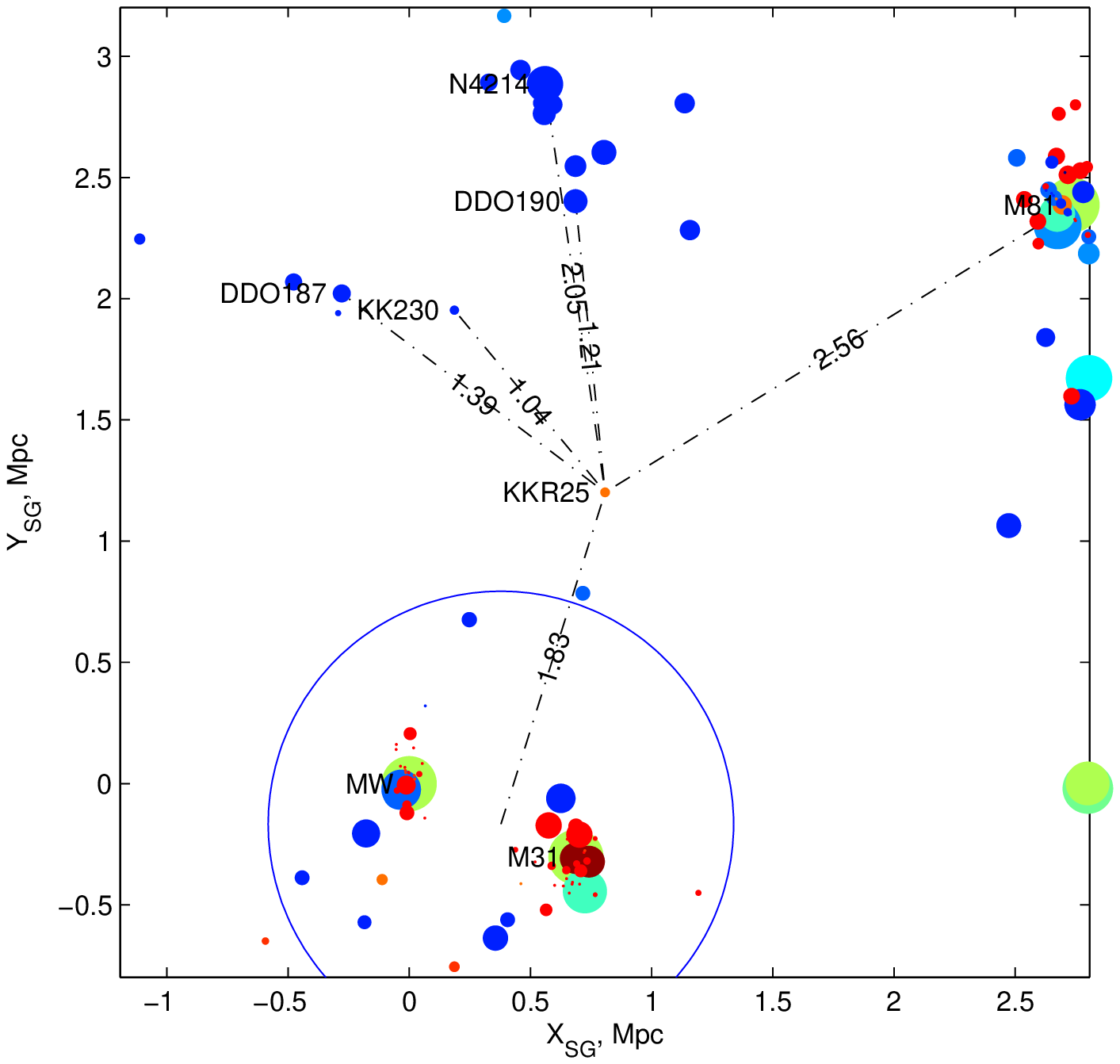} & \includegraphics[height=0.45\textwidth]{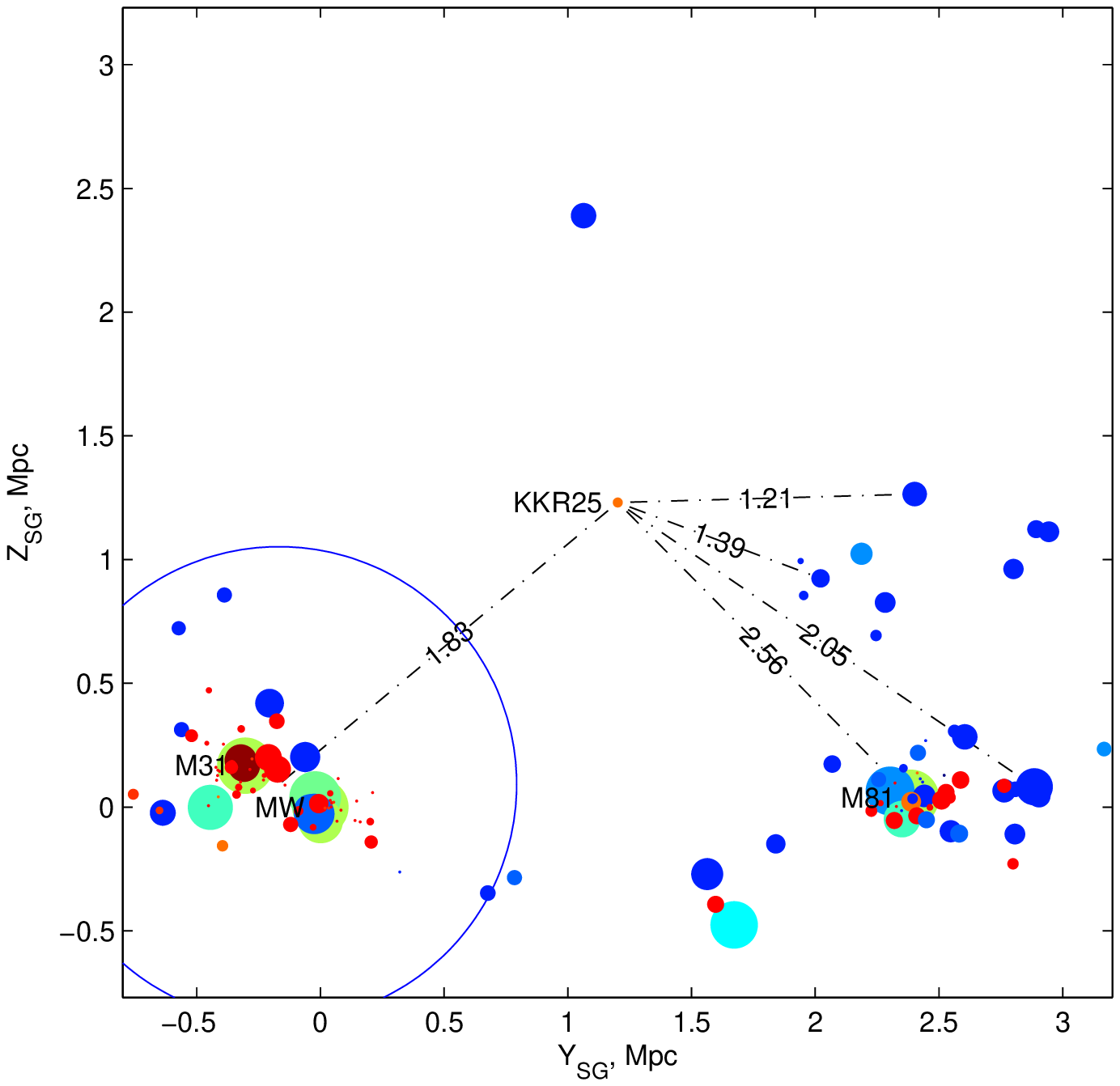}
\end{tabular}
\caption{
The map of galaxies in the SuperGalactic coordinates is centred on KKR\,25.
The left panel presents the projection of galaxies on the SuperGalactic plane, 
while the right panel shows the edge-on view on the `pancake' of galaxies.
The filled circle size is proportional to absolute magnitude of galaxies.
Galaxies are coded by a colour from red for early types ($T=-5$) to blue for late types ($T=10$), according to morphological type in de Vaucouleur's numerical scale \citep{RC3}.
The zero-velocity surface is shown by big blue circle around the Local Group.
Distance to nearby structure is marked by dash-dotted lines with corresponding distance written on it. The brightest galaxies in the volume under consideration are Milky Way (WM), Andromeda galaxy (M31) and M81. They are signed on the figures. KK\,230 is a closest galaxy to KKR\,25.
Three associations of dwarf galaxies \citep{DwarfsAssociations} are shown on left panel. 
DDO\,190 is the brightest member of 14+08 association. DDO\,187 corresponds to `Dregs' association and NGC\,4214 to 14+07. The Local Void occupies the upper half of the right panel just above KKR\,25.
}
\label{f:XYZ}                                                 
\end{figure*}                                                 

The isolated dwarf spheroidal galaxy KKR\,25 was discovered by \citet{KKR1999} during the search of dwarf galaxies in the direction of the Local Void. Follow-up observations with the 100-m radio telescope at Effelsberg \citep{HI2,HI5} have shown \hi\ emission in the object with a radial velocity $V_h=-139.5$ \kms. Direct images of this low surface brightness galaxy were obtained with
the 6-meter telescope of the Special astrophysical observatory (SAO) of the Russian Academy of Sciences \citep{KKR25+HST} and with the Hubble Space Telescope (\textit{HST}). Its colour-magnitude diagram (CMD) shows a red giant branch population and a trace of blue stars. \citet{KKR25+HST} classify the object as transition type galaxy (dIrr/dSph) at the distance of 1.86 Mpc.

On the other hand, the spectral survey of nearby dwarf LSB galaxies with the Russian 6-m telescope
failed to detect an optical velocity of KKR\,25 \citep{NearbyLSB+Vh}.
Moreover, deep radio observations with the Giant Metrewave Radio Telescope (GMRT) \citep{KKR25+GMRT} did not show significant \hi\ emission in the range $-256<V_h<-45$ \kms\ at the level $M_{HI}=0.8\times10^5$ $M_{\sun}$. \citet{KKR25+GMRT} concluded that `the non-detection of \hi\ in KKR\,25 suggests that previous single-dish measurements were affected by confusion with the Galactic emission. Our stringent limits on the \hi\ mass of KKR\,25 indicate that it is a normal dSph galaxy'.

KKR\,25 is one of the most isolated galaxies inside the sphere of 3 Mpc around us. 
It settles at the distance of 1.9 Mpc from the Milky Way and at 1.2 Mpc above the SuperGalactic plane in the front of the Local Void. 
KKR\,25 is far away from the zero-velocity surface of $R_0=0.96\pm0.03$ Mpc \citep{KKMT2009}, which separates the Local Group from the cosmic expansion. 
The Local Group is the nearest massive structure to KKR\,25.
The second close massive group is the M\,81 at the distance of 2.56 Mpc.
There are no galaxies closer than 1 Mpc to KKR\,25 (see Fig.\ \ref{f:XYZ}). 
The nearest neighbour is the dwarf galaxy KK\,230 ($M_B=-9.8$). 
The isolation of KKR\,25 was pointed out by \citet{KKR25+HST}. 
\citet{DwarfsAssociations} note that KKR\,25 is only isolated galaxy on the scale of 3 Mpc from us and `every object in this volume is associated with either a luminous group, an association of dwarfs, or the dregs evaporating association'. 
Two associations 14+08 (around DDO\,190) and `Dregs' (around DDO\,187) stand on the distance of 1.2 and 1.4 Mpc from KKR\,25, respectively.

In spite of its isolation KKR\,25 has no gas and looks like a normal dwarf spheroidal system.
This fact draws our attention, because we expect to find dSph galaxies in dense regions, like groups and clusters of galaxies. 
Obviously, that any kind of interaction with massive galaxy is not suitable to explain properties of KKR\,25. 
This galaxy can play a crucial role in testing of different scenarios of dSph's formation. 

\citet{KKR25+HST} have found a globular cluster candidate in the HST images of the galaxy.
An apparent magnitude of the object $V_T=20.59$ corresponds to $M_V=-5.79$, which is typical for Galactic globular clusters. 
In the framework of the \Halpha{} survey of the Canes Venatici I cloud of galaxies with the Russian 6-m telescope \citet{CVnI+Halpha} also have found a faint \Halpha{} knot on northern side of KKR\,25. 
A measured flux of the knot is $\log F=-14.64$ erg\,cm$^{-2}$\,s$^{-1}$.
These interesting objects were targeted for spectroscopic study with 6-m telescope in current study.

The main parameters of KKR\,25 are presented in the Table~\ref{t:KKR25}.
Coordinates are taken from HyperLEDA\footnote{\url{http://leda.univ-lyon1.fr/}} \citep{HyperLEDA}.
Apparent sizes were published by \citet{KKR1999}. The colour $(V-I)_T$ was measured by \citet{KKR25+HST}.
The central surface brightness $\Sigma_{V}$ was estimated from profiles published by \citet{KKR25+HST}. 
All other values, total magnitude $V_T$, axis ratio, scale length $h$, heliocentric velocity $V_h$ 
and distance modulus $(m-M)_0$, are derived in the current work.
The $V_T$, $(V-I)_T$ and $\Sigma_{V}$ magnitudes are not corrected for Galactic extinction. 

\begin{table}
\caption{Main parameters of KKR\,25.}
\begin{tabular}{@{}lr@{\,\,}ll}
R.A. (J2000)                    & \multicolumn{2}{c}{$16\,13\,47.6$}  & HyperLEDA \\
Dec (J2000)                     & \multicolumn{2}{c}{$+54\,22\,16$}   & HyperLEDA \\
$E(B-V)$, mag                   & 0.008           &                   & \citet{DustMap} \\[3pt]
Size, arcmin                    & \multicolumn{2}{c}{$1.1\times0.65$} & \citet{KKR1999} \\
$h$, arcsec                     & 16.7            & $\pm 1.1$         & this work \\
$b/a$                           & $0.51$          & $\pm0.03$         & this work \\[3pt]
$V_T$, mag                      & $15.52$         & $\pm0.22$         & this work \\
$(V-I)_T$, mag                  & $0.88$          &                   & \citet{KKR25+HST} \\
$\Sigma_{V}$, $\textrm{mag}/\square^{\prime\prime}$     & $23.97$         & $\pm0.03$         & \citet{KKR25+HST} \\[3pt]
$V_h$(stars), \kms              & $-65$           & $\pm15  $         & this work \\
$V_h$(PN), \kms                 & $-79$           & $\pm9   $         & this work \\[3pt]
$(m-M)_0$, mag                  & $26.42$         & $\pm0.07$         & this work \\
Distance, Mpc                   & $1.93$          & $\pm0.07$         & this work \\[3pt]
$V_{LG}$, \kms                  & $128$           &                   & this work \\
$M_V$, mag                      & $-10.93$        &                   & this work \\
$L_V$, $10^6 L_\odot$           & $2.0$           &                   & this work \\
$\Sigma_V$, $L_\odot$\,pc$^{-2}$& $9.6$           &                   & this work
\end{tabular}
\label{t:KKR25}
\end{table}

\section{Direct images}

\subsection{Observations and photometry}

\begin{figure}
\centerline{
\includegraphics[width=0.45\textwidth]{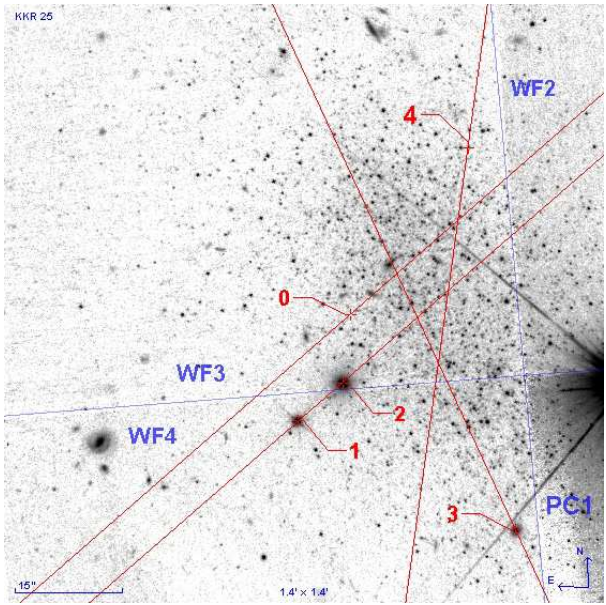}
}
\caption{
\textit{HST} WFPC2 image of KKR\,25 in \textit{F814W} band.
The bounds of planetary (PC1) and wide field cameras (WF2, WF3, WF4) are shown.
Main part of KKR\,25 is arranged in WF3 chip.
A bright star fully saturates the PC1 camera.
The objects for spectroscopic study are numbered.
The long slit positions are overplotted. 
The integrated spectrum of the stellar light was obtained in the slit position `0'.
}
\label{f:image}
\end{figure}

\begin{figure}
\centerline{
\includegraphics[width=0.45\textwidth]{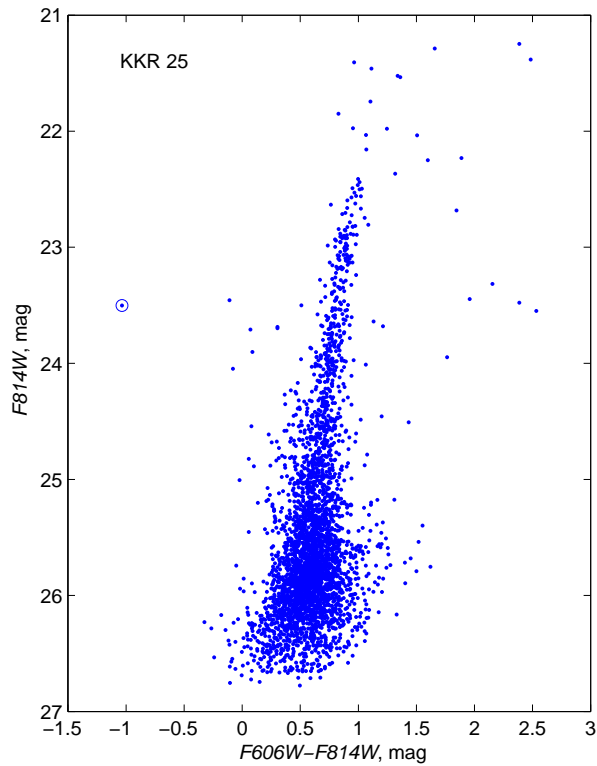}
}
\caption{
The colour-magnitude diagram based on \textit{HST}/WFPC2 image of KKR\,25.
Extremely blue object 4 from our spectroscopic study is shown by open circle.
}
\label{f:CMD}
\end{figure}

KKR\,25 was observed first with WFPC2 on 28 May 2001 as a part of the Hubble Space Telescope snapshot project (proposal 8601, PI: P.\ Seitzer). 
Two exposures were made with filters \textit{F606W} (broadband \textit{V}) and \textit{F814W} (broadband \textit{I}).
The exposure time is 600 seconds in each filter.
Later the galaxy was observed with WFPC2 on 23-25 March 2009 within the ANGST project \citep{ACSTreasury}. 
Deep exposures in \textit{F606W} (4800s) and \textit{F814W} (9600s) were made.
The image of KKR\,25 is shown in Fig.\ \ref{f:image}.
Stellar photometry procedures, artificial star experiments and star formation history (SFH) measurements were made for the both exposure sets. 
The results of the SFH measurements are in good agreement. We have decided to consider only the deeper exposure data in this paper. 
All descriptions below are relate to the project 11986 (PI: J.\ Dalcanton) HST/WFPC2 images. 
The WFPC2 images were obtained from the STScI archive using the standard processing and calibration pipeline. A photometry of resolved stars in the galaxy was made with the \HSTphot{} package \citep{HSTPHOT}, following procedures and recipes indicated in the \HSTphot{} Users Guide\footnote{\url{http://purcell.as.arizona.edu/hstphot/hstphot.ps.gz}}. The data quality images were used to mask bad pixels. Cosmic rays were masked using the \HSTphot{} \textit{crmask} utility. 
Only stars with photometry of good quality (signal-to-noise $S/N\ge5$ in both filters, 
$|\chi|<2.5$, $|\textrm{sharp}|<0.3$ and $\textrm{type}\le2$) were used in the analysis.
Resulting colour-magnitude diagram (CMD) is presented in Fig.\ \ref{f:CMD}.

\begin{figure}
\centerline{
\includegraphics[width=0.45\textwidth]{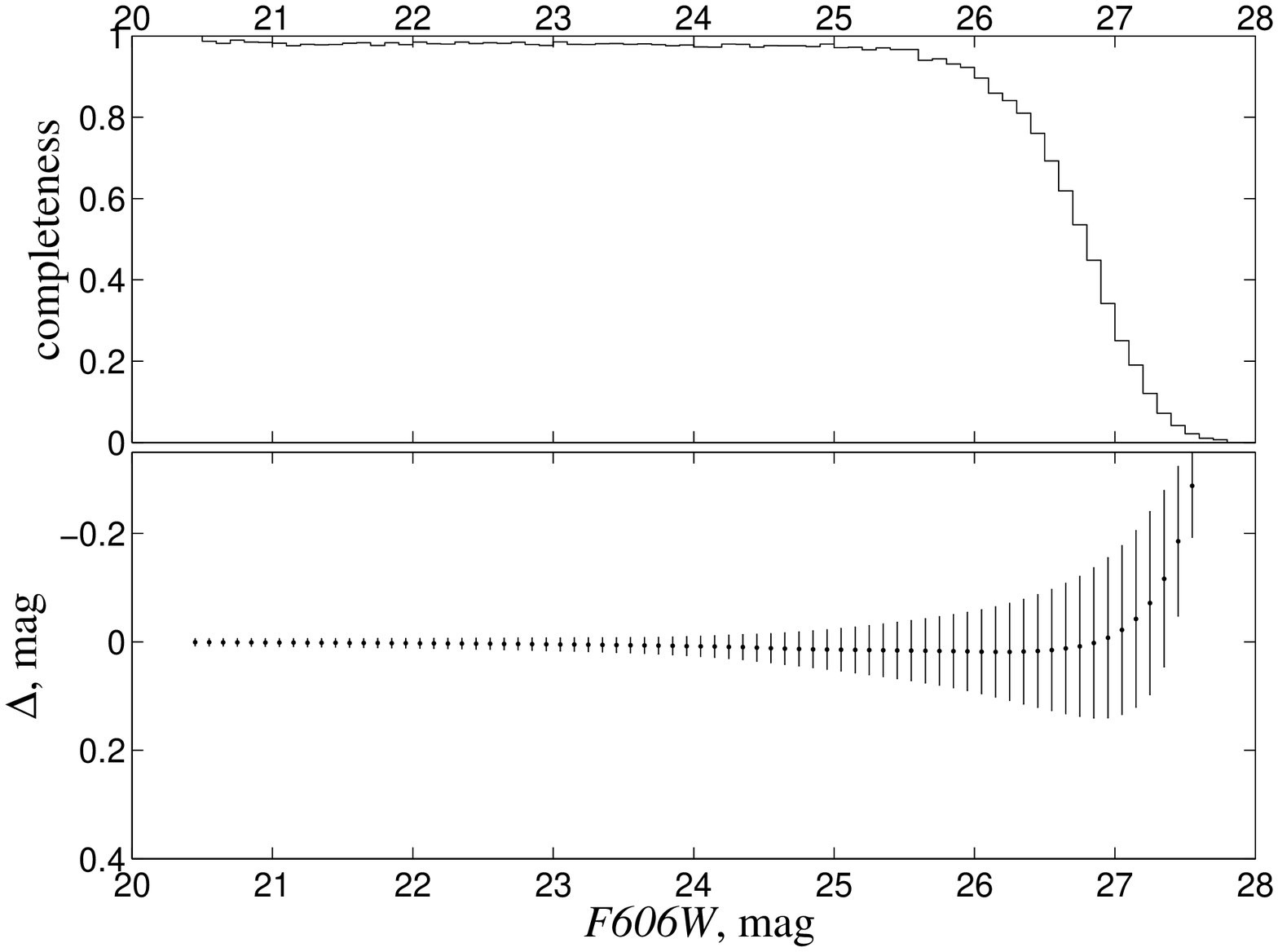}
}
\centerline{
\includegraphics[width=0.45\textwidth]{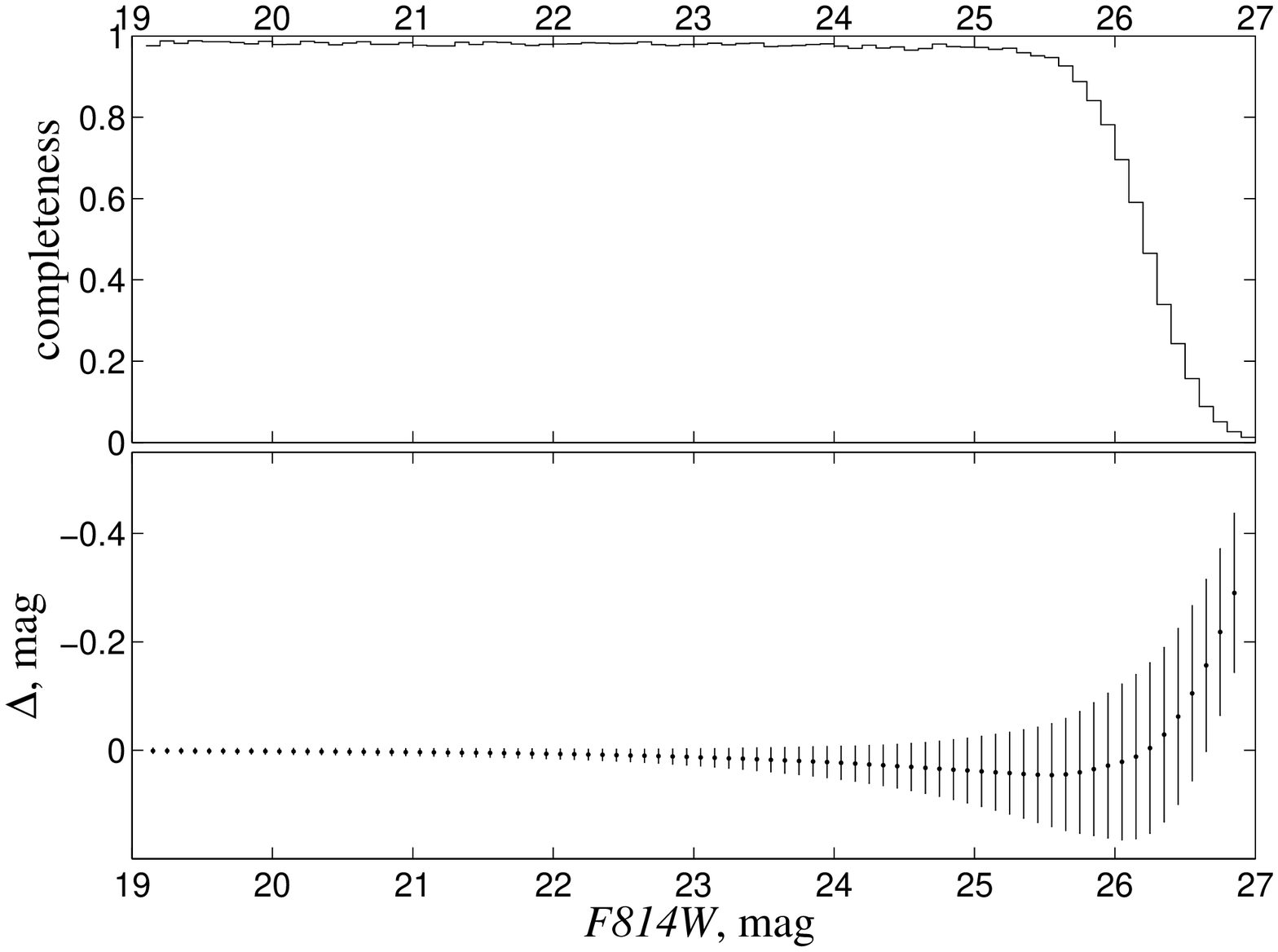}
}
\caption{
Photometric errors and completeness of KKR\,25 in ANGST project.
The photometry statistics were obtained using artificial star tests.
The input, `true' magnitude of artificial stars in \textit{F606W} and \textit{F814W} filters are shown on the abscissa.
The completeness panel represents a fraction of detected artificial stars as function of input magnitude.
The dispersion and bias of difference between `observed' and `true' magnitudes ($\Delta=m_{\rm obs}-m_{\rm true}$) 
are shown on bottom panels for both passbands.
}
\label{f:uncert}
\end{figure}

Artificial star tests were performed using the same reduction procedures 
to estimate photometric errors correctly.
We have created a large library of artificial stars distributed in the same
range of stellar magnitudes and colours as the original measured stars. 
Spatial distribution of the artificial stars also were resembling the original one so that
the recovered photometry is adequately sampled.
The photometric errors and completeness are represented in Fig.\ \ref{f:uncert}.

\subsection{Reddening and background contamination} 

KKR\,25 is situated at the rather high Galactic latitude and Galactic extinction
is not large for this galaxy: $A_V=0.028$ and $A_I=0.016$\,mag according
to \citet{DustMap}. As can be seen from the colour-magnitude diagram of KKR\,25, 
the field is not heavily contaminated by background stars. In order to estimate 
a number of foreground/background stars in the CMD, we have used \textsc{trilegal} 
\citep{TRILEGAL} to simulate star counts in our Galaxy.  
\textsc{trilegal} simulates CMDs in the WFPC2 instrumental system taking into
account the components of thin and thick Galactic disks, the halo and the bulge
of our Galaxy. The photometric errors, the saturation and the incompleteness
were taken into account in the modelling. These models confirm
a negligible contamination by foreground stars.
An expected number of foreground stars in the CMD is about
43$\pm$6 up to the photometric limit $I=27.0$\,mag.

\subsection{Distance determination}

A photometric distance to KKR\,25 was firstly estimated by \citet{KKR25+HST} using
tip of of the red giant branch (TRGB) distance indicator. They have obtained a true
distance modulus $(m-M)_0 = 26.35\pm0.14$ mag and a respective distance $D = 1.86\pm0.12$ Mpc. 
TRGB distances for several galaxies in M\,81 group were recently measured also by
\citet{ACSTreasury} within their ANGST project. 
However, the older dataset was used for their KKR\,25 distance measurement in this paper.
They determined a distance modulus of KKR\,25 $(m-M)_0=26.43$.

The recent deep HST/WFPC2 observations and accurate \HSTphot{} photometry allow us to estimate a TRGB distance with better uncertainties. 
TRGB method was also considerably improved recently. We have determined a photometric TRGB distance with
our \textsc{trgbtool} program, which uses a maximum-likelihood algorithm
to determine the magnitude of tip of the red giant branch from the stellar
luminosity function \citep{TRGB1}. The estimated value of TRGB is equal to
$F814W = 22.41\pm0.07$ mag in the WFPC2 instrumental system. 
The uncertainty of the TRGB value estimation is dominated by poor statistics in 
the considered region of the colour-magnitude diagram. 
The calibration of the TRGB distance indicator was recently improved \citep{TRGB2}. 
Colour dependence of an absolute magnitude of
TRGB and zero-point issues in \textit{HST}/ACS and WFPC2 were addressed in the paper. Using 
this calibration, we have obtained the true distance modulus for KKR\,25 
$(m-M)_0 = 26.42\pm0.07$ mag and respective distance $D = 1.93\pm0.07$ Mpc. 
Both internal and calibration errors were taken into account in the error budget.
The precision of the Galactic extinction value (16 per cent on the Schlegel's maps) was also accounted.
The new value is in good agreement with all the previous distance measurements.

\subsection{Star formation history}
\label{sfh}

The colour-magnitude diagram of KKR\,25 (see Fig.\ \ref{f:CMD}) shows very tight
and clean red giant branch, in the lower part of which, near the photometric
limit (F814W $\sim$ 25.5--26.5), we can see signs of a red clump. Judging from the
CMD, we can expect low abundance and low age spread for the oldest stars in 
the galaxy. The extremely blue object in the CMD was classified by us as a planetary nebula belonging to KKR\,25. 
The spectroscopic study of the PN is giving below. 

First a star formation rate dependence from an age for KKR\,25 was measured by \citet{Weisz+2011} within their sample of 60 nearby galaxies.
We determined detailed star formation history of KKR\,25 from its CMD using our StarProbe package \citep{mm04}. 
This program fits the observed photometric distribution 
of stars in the colour-magnitude diagram to a positive linear combination
of synthetic diagrams of single stellar populations (SSPs, single age
and single metallicity). The algorithm and the package are described in
details in the paper of \citet{mm04}, and the package application to star 
formation history determination for some nearby galaxies are described in
our work \citep{m2010}. The observed data were binned into Hess diagram, 
giving number of stars in cells of the CMD. 
The size of the cells is 0.05 mag in luminosity and colour.
The colour and magnitude limits of the constructed Hess diagram are 
$-0.3 < F606W-F814W < 2.5$, $21.0 < F814W < 26.8$.
Synthetic Hess diagrams 
were constructed from theoretical stellar isochrones and initial mass function (IMF).
We used
the Padova2000 set of theoretical isochrones \citep{girardi00}, and a \citet{salpeter55} IMF.
We create a grid of the isochrones interpolated by age to fill gaps between the
original isochrones. The original metallicity set was not altered.
The distance was taken from the present paper
(see above) and the Galactic extinction is from \citet{DustMap}.
A binary fraction was assumed to be 30 per cents. 
The mass function of individual
stars and the main component of a binary system is
supposed to be the same. The mass distribution for
second component was taken to be flat in the range
0.7 to 1.0 of the main component mass.

\begin{figure}
\centerline{
\includegraphics[width=0.45\textwidth]{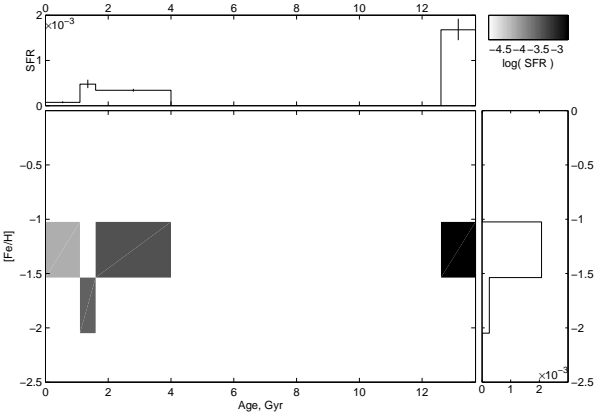}
}
\caption{
The star formation history of the dwarf galaxy KKR\,25.
The top panel shows the star formation rate (SFR) ($M_\odot$/yr) against the
age of the stellar populations.
The bottom panel represents the metallicity of stellar content as function of
age. The grayscale colour corresponds to the strength of SFR for given age and
metallicity.
\label{f:cmd_sfh}
}
\end{figure}

The synthetic diagrams were altered by the same
incompleteness and crowding effects, and photometric systematics
as those determined for the observations using artificial stars
experiments. The synthetic diagrams covers
all the range of ages (from 0\,Myr to 13.7\,Gyr) and metallicities
(from $Z = 0.0001$ to $Z = 0.03$). The best fitting combination of synthetic CMDs is a maximum-likelihood solution
taking into account the Poisson noise of star counts in the cells of Hess diagram.
The resulting star formation history (SFH) is shown in
the Fig.\ \ref{f:cmd_sfh}. The 1\,$\sigma$ error of each SSP is derived from 
analysis of the likelihood function.

The star formation histories from our work and \citet{Weisz+2011} are in general good agreement.
The main difference is in duration of star formation episodes.
\citet{Weisz+2011} used 6 fixed periods of star formation for all 60 galaxies, 
while in our measurement we have sophisticated algorithm of star formation periods fitting.
Firstly, we divided the galaxy lifetime into quite small steps 
(0.5 Gyr between 0 and 2 Gyr and 2 Gyr between 2 and 13.7 Gyr). 
In the resulting SFH we selected 5 episodes of star formation with constant SFR. 
A variation of bounds between these episodes of star formation allows
to determine the best fitting of the models to the CMD.

According to our measurements, a main star formation event in KKR\,25 has occurred 12.6 -- 13.7 Gyr ago 
with a mean star formation rate (SFR) $1.7\pm0.2\times10^{-3}$ \msol{} yr$^{-1}$. 
It is the total SFR over the whole galaxy. 
A metallicity range for the stars formed during this event is about 
$\feh=[-1.6:-1]$\,dex. This initial burst accounts for 62 per cents 
of the total mass of formed stars.

A quiescence period has appearing from about 4 to 12.6 Gyr ago.
There are indications of intermediate age star formation in KKR\,25 between 
1 and 4\,Gyr with no significant signs of metal enrichment for these stars. A
star formation rate is lower in this period and is equal
$3.6\times10^{-4}$ \msol{} yr$^{-1}$. 
The measured star formation rate is very low for the
recent 1 Gyr in this dwarf galaxy: $0.7\pm0.3\times10^{-4}$ \msol{} yr$^{-1}$.
We found a total mass of the formed stars during KKR\,25 lifetime to be
$3.0\pm0.3\times10^6$ \msol{}.

\section{Structure of the galaxy}

Total and surface photometry of KKR\,25 are difficult tasks because of a bright foreground
star at 20 arcsec from the centre of the galaxy.
However, the \HSTphot{} stellar photometry is reliable and affected by bright star much less. 
Moreover, this influence can be taken into account quantitatively.

\begin{figure}
\centerline{
\includegraphics[width=0.45\textwidth]{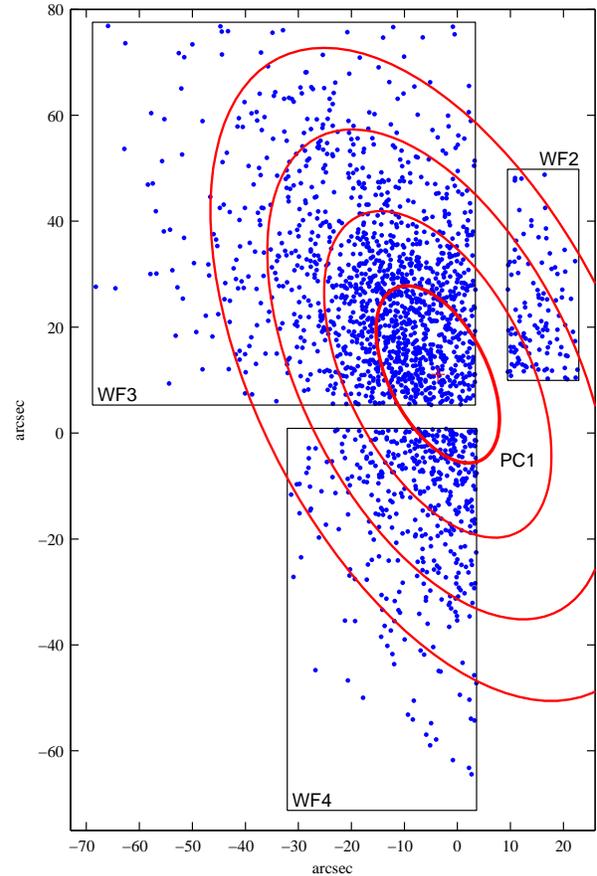}
}
\caption{
Distribution of the selected stars ($F814W<26$) of KKR\,25 in the frame of WFPC2.
The boxes represent the considered areas of the cameras WF2, WF3 and WF4.
The PC1 camera affected by bright star and was not included in the analysis.
The thick ellipse shows the size of central region of constant stellar density.
The set of thin ellipses corresponds to 2, 3 and 4 times of the exponential scale length.
The cross indicates the centre of the galaxy.
\label{f:xyfit}
}
\end{figure}

We used the stellar photometry of the deep WFPC2 images to study a distribution of the stars in the galaxy.
Our approach based on the maximum likelihood fit that was described by \citet{MdJR2008}.
The likelihood function of stellar distribution in focal plane is defined as
\begin{equation}
\mathcal{L}=\prod_i^N l_i(p_1,p_2,\dots,p_k),
\end{equation}
where $l_i(p_1,p_2,\dots,p_k)$ is probability to find 
a star $i$ given by the set of parameters $p_1,p_2,\dots,p_k$ with total number of stars $N$. 
This probability is determined by the surface density distribution of foreground and galactic 
stars, as well as the photometric completeness. 
Since KKR\,25 is located at the high Galactic latitude ($b=+44.4$) and the contamination of 
CMD by foreground stars is very insignificant we can neglect contribution of foreground objects.
For the analysis we have selected only stars with $F814W<26$, which corresponds to completeness level of about 80 per~cent.
We have tested the behaviour of completeness over the field of view.
We did not find valuable variations in the body of the galaxy except for regions near brightest foreground stars. 
For instance, the bright foreground star affects the PC1 and partially WF2 cameras which is seen in Fig.\ \ref{f:image}.
We excluded these regions from consideration. 
Distribution of stars in the areas under consideration is shown in the Fig.\ \ref{f:xyfit}.
The probability can be expressed in the form
\begin{eqnarray}
&& l_i(p_1,p_2,\dots,p_k) = \Sigma(x_i,y_i|p_1,p_2,\dots,p_k) \\
&& N = \int\!\!\!\!\int_D \Sigma(x,y|p_1,p_2,\dots,p_k)\,\mathrm{d}x\,\mathrm{d}y
\end{eqnarray}
where $\Sigma$ is the model of surface density distribution with $k$ parameters $p_1,p_2,\dots,p_k$
and $\mathrm{D}$ is consideration domain. 

We tested a set of models of surface density distribution:
the exponential, exponential with central depression, \citet{King}, \citet{Plummer} 
and \citet{Sersic} profiles.

The WFPC2 coordinates of 1859 selected stars were corrected for the distortion and 
were transformed to the reference chip using \textit{metric} procedure from \textsc{stsdas.hst\_calib.wfpc} package.
For each model we derived a centre, position angle and axes ratio of the galaxy (4 parameters) 
as well as scale length parameters of the model (from 1 to 3 parameters). 
The central surface density is derived as normalisation coefficient to the total number of stars.

We used the Akaike Information Criterion $\textrm{AIC}=-2\ln\mathcal{L}+2k$ \citep{AIC} and
the Bayesian Information Criterion $\textrm{BIC}=-2\ln\mathcal{L}+k\ln N$ \citep{BIC},
where $N$ is the sample size, i.e.\ the number of selected stars in our case.
These criteria regulate the balance between the improvement of model by additional parameters and 
the number of parameters needed to achieve this improvement. 
One should prefer a model with a minimal value of the criterion.
An absolute value of the criterion is not informative, while the difference 
between the AIC or BIC values for two fits provides an estimate of evidence
for one model against another.
The BIC difference $0<\Delta_{\rm BIC}<2$ is not worth more than a bare mention, 
$2<\Delta_{\rm BIC}<6$ represents positive evidence, 
$6<\Delta_{\rm BIC}<10$ means strong evidence
and $\Delta_{\rm BIC}>10$ very strong evidence \citep{KR1995}.
There are similar guidelines for $\Delta_{\rm AIC}$ values \citep{BA2004}.
More details on a model selection can be found, for instance, in \citet{Liddle2007}.

\begin{table}
\caption{
A comparison of the different models of stellar population distribution of KKR\,25.
$\Delta_{\rm AIC}$ and $\Delta_{\rm BIC}$ are computed respectively to the model with minimal value of the criterion. 
\label{t:sdfit}}
\begin{tabular}{lrrrr}
\hline
Model       & $k$   & $\ln\mathcal{L}$  & $\Delta_{\rm AIC}$  & $\Delta_{\rm BIC}$  \\
\hline
exp         &   5   & $-9963.4$         & 43.6     &  38.1  \\
exp 1       &   6   & $-9940.6$         & \underline{0.0}     &   \underline{0.0}  \\
exp 2       &   7   & $-9940.3$         &  1.5     &   7.0  \\
Plummer     &   5   & $-9956.0$         & 28.9     &  23.4  \\
Sersic      &   6   & $-9949.2$         & 17.2     &  17.2  \\
King        &   6   & $-9946.4$         & 11.7     &  11.7  \\
\hline
\end{tabular}
\end{table}

A comparison of different models is presented in Table~\ref{t:sdfit}.
The `exp 1' and `exp 2' code the models of exponential profile with central depression.
These models differ from simple exponent within specific radius near the centre of the galaxy. 
The `exp 1' corresponds to the model with a constant density in the centre,
the `exp 2' is an exponential decay in the central part of the galaxy.
Taking into account only likelihood value, we can conclude that
the pure exponential profile is the worst case of all.
However, the simple modification of the exponential model with depression in the centre 
improves the fit significantly, and these models leave behind all other models under consideration.
The AIC and BIC clearly select exponential profile with the constant surface density
in the central region (exp 1) as the best fit model for the given distribution of the stars.

\begin{figure}
\centerline{
\includegraphics[width=0.45\textwidth]{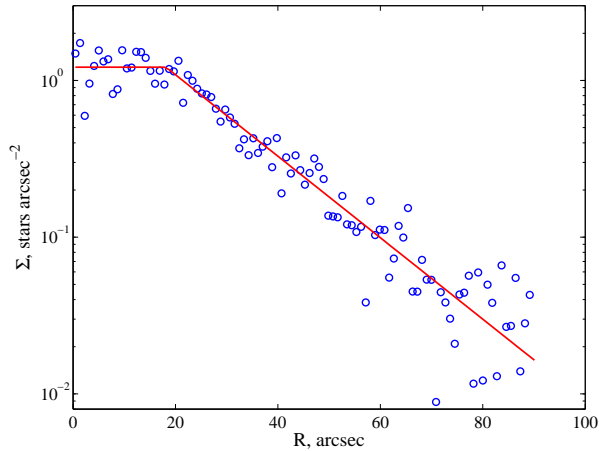}
}
\caption{
Surface density profile of the selected stars in KKR\,25.
The stellar density is measured in concentric elliptical annuli from the centre.
The solid line represents the best fit model, the exponential profile with the
depression in the central region.
The parameters of the model were determined using maximum likelihood method.
\label{f:sdfit}
}
\end{figure}

\begin{table}
\caption{
Parameters of the `exp 1' model for different star selections. 
$h$ is an exponential scale length,
$b/a$ is an axes ratio of the stellar distribution
and
$R$ is the central depression radius.
\label{t:sdbest}}
\begin{tabular}{lr@{ }lr@{ }lr@{ }l}
\hline                                       
                        &\multicolumn{2}{c}{all}       &\multicolumn{2}{c}{RGB}         &\multicolumn{2}{c}{RC}               \\
                        &\multicolumn{2}{c}{$F814W<26$}&\multicolumn{2}{c}{$F814W<25.5$}&\multicolumn{2}{c}{$25.5<F814W<26$}  \\
\hline                                       
$h$, $^{\prime\prime}$  &   16.7 & $_{-1.0}^{+1.1}$    &   18.4 & $_{-1.4}^{+1.7}$      &   15.7 & $_{-1.3}^{+1.4}$           \\[3pt]
$h$, pc                 &   156  & $_{-11}^{+12}$      &   172  & $_{-14}^{+17}$        &   147  & $_{-13}^{+14}$             \\[3pt]
$b/a$                   &   0.51 & $_{-0.03}^{+0.03}$  &   0.46 & $_{-0.04}^{+0.04}$    &   0.55 & $_{-0.04}^{+0.05}$         \\[3pt]
$R$, $^{\prime\prime}$  &   18.2 & $_{-3.1}^{+2.3}$    &   12.4 & $_{-6.4}^{+4.3}$      &   20.5 & $_{-2.8}^{+2.8}$           \\[3pt]
$R$, pc                 &   170  & $_{-30}^{+22}$      &   116  & $_{-60}^{+40}$        &   192  & $_{-27}^{+27}$             \\
\hline
\end{tabular}
\end{table}

Stellar density profile of KKR\,25 is shown in Fig.\ \ref{f:sdfit}.
Structural parameters of the best model are summarized in the Table~\ref{t:sdbest}.
The galaxy shows the exponential profile with axial ratio of 1:2 and 
flat distribution of the surface brightness near the centre.
The angular size of the central depression roughly equals to the exponential scale length.
We have estimated an integrated magnitude of the galaxy from the derived model parameters (see Table~\ref{t:sdbest}, `all')
and the central surface density $\Sigma_V=23.97$ and $\Sigma_I=23.09$ mag,
which were taken from our previous photometry results \citep{KKR25+HST}.
We obtained the total magnitudes of KKR\,25 $V_T=15.52\pm0.22$ and $I_T=14.64\pm0.22$. 
The new $V_T$ value is 0.4 mag brighter than $V_T=15.9$ obtained by \citet{KKR25+HST}.
The difference can be explained by difficulties of photometry caused by nearby bright star 
and the small field of view where the galaxy occupies whole WF3 chip of WFPC2.

There is an indication on slightly different spatial distribution of red giant branch and red clump stars.
These populations are indicated as RGB and RC in the Table~\ref{t:sdbest}, respectively.
Judging from the scale length, the RC population is slightly more centrally concentrated than RGB stars.
But the difference appears on level of two sigma.

\section{Spectroscopy}

\subsection{Observations and data reduction}

We carried out several sets of long-slit observations of 
a globular cluster candidate, bright objects and \Halpha{} source in KKR\,25.
The spectroscopic data were obtained with the multi-mode focal reducer 
SCORPIO\footnote{\url{http://www.sao.ru/hq/lsfvo/devices/scorpio/scorpio.html}}
\citep{scorpio} in the prime focus of the Russian 6-m telescope.

We used the grism VPHG550G (550 lines\,mm$^{-1}$) and 2k$\times$2k CCD EEV42--40 detector.
This combination together with the slit of $6\arcmin\times1\arcsec$ provides
a spectral resolution $\sim\!10$\AA{} in the range of 3500--7200\AA. 
A typical dispersion was 2.1\AA{} per pixel.
The frames were binned by 2 pixels along the slit direction, 
resulting in a spatial scale of 0.357\,arcsec per pixel.
For wavelength calibration we used He-Ne-Ar lamp. 

The journal of observations is presented in Table~\ref{t:obslog}.
Orientations of the slit are shown on the direct WFPC2 image of the galaxy in Fig.\ \ref{f:image}. 

\begin{table}
\caption{The journal of spectroscopic observations.}
\label{t:obslog}
\scriptsize
\begin{tabular}{lcllc}
\hline
Object         & Date            & Exposure (s)  & Seeing (\arcsec) \\ 
\hline
KKR\,25 slit pos.~1--2 & 08/02/05 & $\phantom{0}900 \times  2$ & 2.0 \\
KKR\,25 slit pos.~3    & 09/02/05 & $\phantom{0}900 \times  3$ & 1.6 \\
KKR\,25 slit pos.~0    & 10/02/05 & $1200 \times 3, 900$ & 2.5 \\
KKR\,25 slit pos.~4    & 30/05/09 & $1200 \times 3, 900$ & 2.3 \\
KKR\,25                & 07/08/11 & $1200 \times 3$      & 1.5 \\
\noalign{\smallskip}
\noalign{Radial velocity standard stars}
\noalign{\smallskip}
BD\,+23\,992 & 08/02/05 & $10         $ &  2.0 \\
             & 09/02/05 & $10 \times 2$ &  2.0 \\
             & 10/02/05 & $10 \times 2$ &  2.5 \\
BD\,+23\,680 & 08/02/05 & $10 \times 2$ &  2.0 \\
BD\,+23\,655 & 09/02/05 & $30 \times 2$ &  2.0 \\
BD\,+23\,751 & 10/02/05 & $10 \times 2$ &  2.5 \\
BD\,+23\,708 & 10/02/05 & $20         $ &  2.5 \\
\noalign{\smallskip}
\noalign{Spectroscopic standard stars}
\noalign{\smallskip}
SA\,95--42   & 08/02/05 & $60         $ &  2.0 \\
BD\,+75\,325 & 09/02/05 & $30 \times 2$ &  2.0 \\
             & 10/02/05 & $30 \times 2$ &  2.2 \\
BD\,+33\,2642& 30/05/09 & $60         $ &  2.6 \\
LDS\,749\,B  & 07/08/11 & $30 \times 2$ &  1.5 \\
\hline 
\end{tabular}
\end{table}

The standard data reduction and analysis were performed using 
the European Southern Observatory Munich Image Data Analysis System\footnote{\url{http://www.eso.org/sci/software/esomidas/}}
(\textsc{eso-midas}) \citep{banse} and 
the Image Reduction and Analysis Facility (\textsc{iraf}) software\footnote{\url{http://iraf.noao.edu/}}. 
The dispersion solution determines an accuracy of the wavelength calibration $\sim0.14$\AA.
The wavelength zero point can be shifted up to 2 pixels during the night. 
This effect was corrected using the night-sky lines in the dispersion-corrected spectra. 
An extraction of the spectra was made using the \textsc{iraf} procedure \textit{apsum}.
After the wavelength calibration and sky subtraction, the spectra were
corrected for atmospheric extinction and flux-calibrated using the observed
spectrophotometric standard stars \citep{oke}. 
Finally, all one-dimensional spectra of each object were summed to increase the S/N ratio. 
We observed bright radial velocity standard stars \citep{bbf}
at the end of each night to obtain heliocentric radial velocities of the 
program objects using the method of \citet{td}, and to obtain
the line-spread function (LSF) of the spectrograph.

\subsection{Integrated stellar light of KKR\,25}

\begin{figure}
\includegraphics[width=0.34\textwidth,angle=-90]{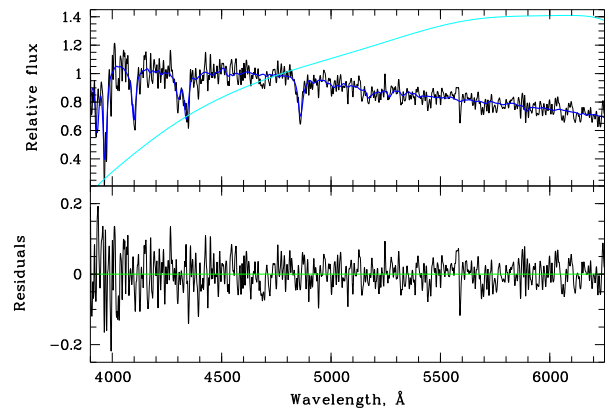}
\caption{
Top: Integrated spectrum of the stellar light of KKR\,25 (black) 
in comparison with a composite model (dark-blue).
The fitting is produced with the \citet{vaz10} SSP model and the Miles stellar library.
The result of the division of the normalized model and object spectra is demonstrated in light-blue.
Bottom: The difference between normalized observed and model spectra.
The zero line is green.}
\label{f:ulyss}
\end{figure} 
 
The spectra of integrated light of the KKR\,25 were obtained during the observations on 10/02/2005 
(Table~\ref{t:obslog}). 
We combined these spectra with the data received in 2009 and 2011 years.
The S/N per wavelength measured in the total integrated light spectrum  at 
$\lambda\sim5500$\AA{} is 54,  at $\lambda\sim4000$\AA{} is 15, 
and goes down at $\lambda \ge 5577$\AA{} due to the bright emission sky lines. 
The resulting spectrum is demonstrated in black in Fig.\ \ref{f:ulyss} (top).

The \textsc{steckmap} program \citep{ocvirk06a, ocvirk06b} with the \textsc{PEGASE.HR} 
model grids \citep{pegase} and the 
\textsc{ULySS}\footnote{\url{http://ulyss.univ-lyon1.fr}}  program
with the \textsc{PEGASE.HR} model grid and the Elodie \citep{prugniel07} and Miles \citep{san06}
stellar libraries\footnote{http://ulyss.univ-lyon1.fr/models.html}, 
and Vazdekis \citep{vaz10} SSP model with the Miles stellar library
were employed to analyze stellar populations in  KKR\,25. 
We use \citet{salpeter55} IMF in all these cases.
Comparison the fitting results for the three sets of models allows to 
claim that $\sim 70$ -- 100 per cent of the light fraction 
in KKR\,25 is as old as 2 -- 14 Gyr and has a low
mean metallicity $\feh \sim-1.75 \pm 0.15$~dex.
Unfortunately, because of low surface brightness of KKR\,25 
the night sky lines affect the resulting spectrum, especially in red part.
Thus we can not realiably estimate a contribution and properties of 
the second stellar component with an age 1 -- 2 Gyr.

The one of three fittings is illustrated in Fig.\ \ref{f:ulyss}, 
where a two-component model (\citet{vaz10} SSP model with the Miles stellar library) is shown in dark-blue.
This model is completely composed of old metal-poor 
(age $\sim18$ Gyr, $\feh=-1.68$ dex) population.
The difference between the normalized object and model spectra 
(Fig.\ \ref{f:ulyss}, bottom) is less equal $10$ per cent. 
Broad absorption-line features are noticeable near 4227, 4455, 
$\sim$4700, 5036, $\sim$5200 which were not fitted well by the models. 
They may be composed of  TiO,  Ca\,\textsc{i}, Mn\,\textsc{i} and Mg\,\textsc{i} lines. The relative intensity of such lines may 
indicate substantial $\alpha$ element enhancement, 
and the presence of dust in the interstellar medium blown out by previous generations of asymptotic-giant branch stars \citep{mboyer08}. 
The obtained \feh{} is in good agreement with the stellar population CMD study presented in Sect.\,2.4.
We determined a radial velocity of the stellar component in KKR\,25 is 
$V_h = -65 \pm 15$ \kms.

\subsection{Bright objects in the field of KKR\,25}

\begin{table}
\caption{
List of spectroscopically observed objects.
The columns are: 
(1) object number in Fig.\ \ref{f:image};
(2), (3) V-band magnitude and $V-I$ colour (\textsc{hstphot} in the case of PN and aperture photometry in all other cases);
(4) heliocentric radial velocity in \kms{}; 
(5) classification.}
\label{t:list}
\scriptsize
\begin{tabular}{l@{ }c@{ }r@{ }r@{ }r@{ }r}
\hline
\multicolumn{1}{c}{N}  &     
\multicolumn{1}{c}{R.A. (2000) Dec.}  & 
\multicolumn{1}{c}{$V$} & 
\multicolumn{1}{c}{$V-I$} & 
\multicolumn{1}{c}{$V_h$} & 
\multicolumn{1}{c}{Note}  \\ 
\hline
 \noalign{\smallskip}   
1       &  161350.0+542201  & $21.6\pm0.13$           & $3.2\pm0.10$           & $ -350 \pm 52$   & star  \\
2       &  161349.3+542206  & $20.5\pm0.12$           & $1.7\pm0.10$           & $ 100431:$       & S0    \\
3       &  161346.5+542145  & $21.3\pm0.15$           & $1.0\pm0.25$           & $ 224844:$       & QSO   \\
 \noalign{\smallskip}
4       &  161347.2+542239  & $22.11\pm0.11$          &  $-1.54\pm0.14$        &  $ -79 \pm 9$    & PN    \\
\hline 
\end{tabular}
\end{table}

The diffuse globular cluster (GC) candidates (2, 3) and a very red star (1) 
were selected for spectroscopic survey using the snapshot \textit{HST} images (see Fig.\ \ref{f:image}). 
Spectra of these objects were obtained with the 6-m telescope in 2005 (Table~\ref{t:obslog}) in the framework
of the project `Searches for GCs in nearby dwarf galaxies' (PI: T.H. Puzia). 
The targets are indicated by numbers in Fig.\ \ref{f:image}.
The candidate found by \citet{KKR25+HST} has number 2 in our list.
The results of the survey are summarised in the Table \ref{t:list}.

\begin{figure}
\centerline{
\includegraphics[height=0.45\textwidth,angle=-90,bb=213 70 523 736,clip]{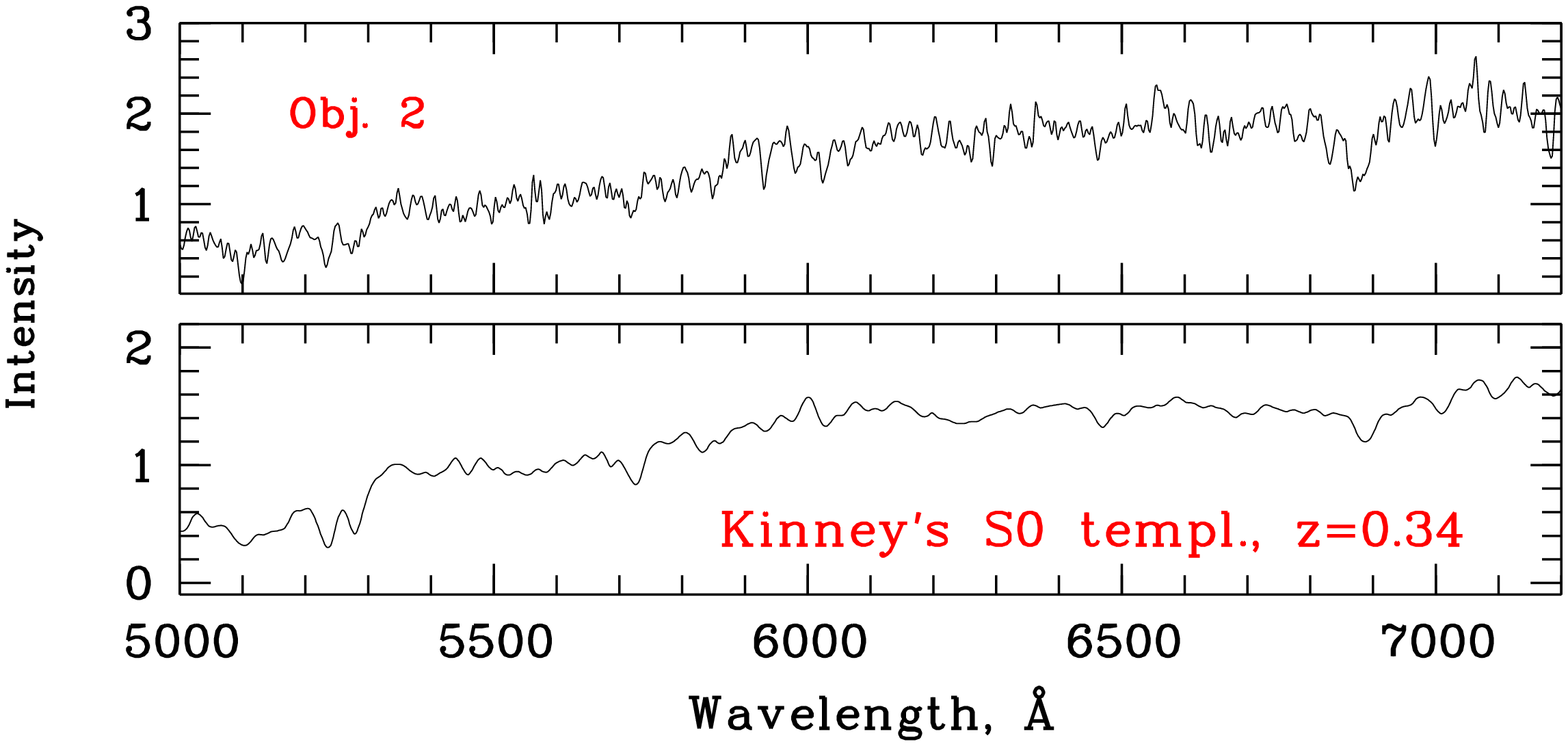}
}
\centerline{
\includegraphics[height=0.45\textwidth,angle=-90,bb=213 70 523 736,clip]{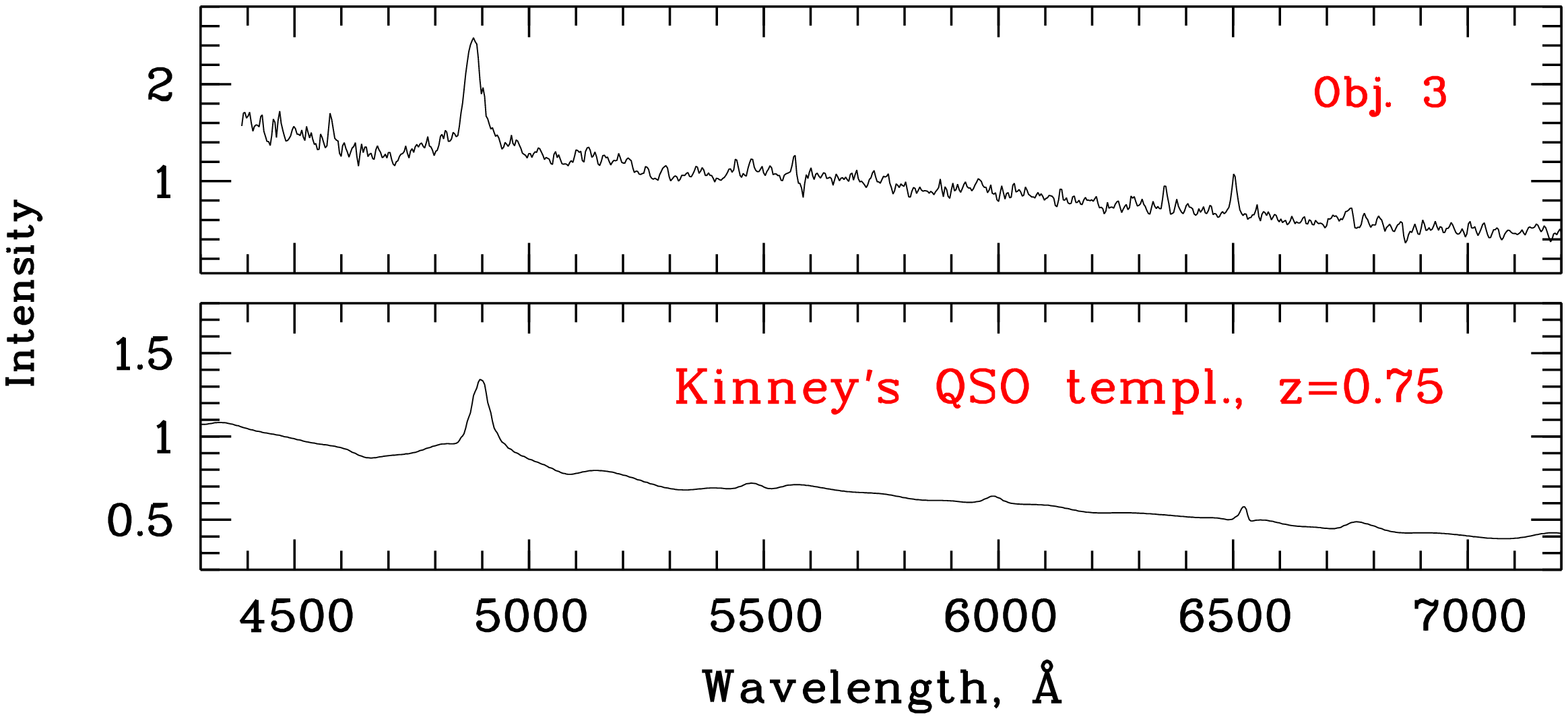} 
}
\caption{
Spectra of background objects (2 and 3, Fig.\ \ref{f:image}) obtained at the 6-m telescope. 
\citet{kinney} templates are shown for comparison.
}
\label{f:spectra}
\end{figure}

\begin{figure}
\centerline{
\includegraphics[width=0.34\textwidth,angle=-90]{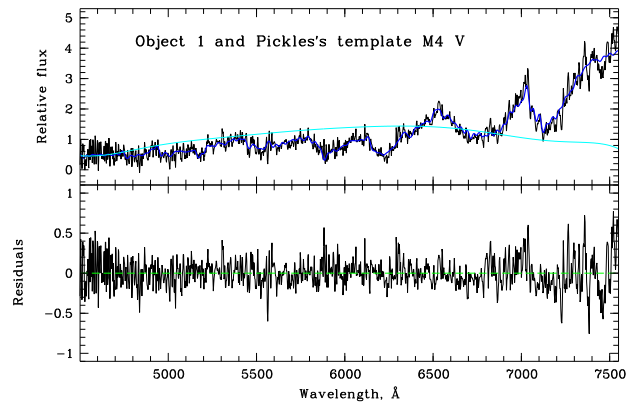}
}
\caption{
Spectrum of a bright red star near KKR\,25.
Top panel shows the observations (black) in comparison with normalized template \citep{pickles} of M4V star (dark blue).
The light-blue line demonstrates a multiplicative term applied
to the original spectrum to bring it into the correspondence to the stellar template.
The lower panel shows the fitting residuals.
}
\label{f:obj1}
\end{figure}

The spectra of GC candidates are shown in Fig.\ \ref{f:spectra}. 
We compared them with the templates of \citet{kinney}.
The object selected by \citet{KKR25+HST} appear to be a S0 galaxy with redshift $z\sim0.335$.
Another bright target is a quasar at $z \sim 0.75$.

The spectrum of the object 1 revealed the nature of this star as a M-type dwarf (see Fig.\ \ref{f:obj1}).
Therefore it does not belong to KKR\,25.

As a result of our survey we did not confirm the presence of globular clusters in KKR\,25.

\subsection{\Halpha{} object}

A faint \Halpha{} emission at the northern side of KKR\,25 was discovered by \citet{CVnI+Halpha} 
in their \Halpha{} survey of galaxies in the Canes Venatici I cloud.
However, WFPC2 image does not show any extended object in the corresponding area of the galaxy.
We identified \Halpha{} object with unusual extremely blue star
which is indicated with number 4 in the direct image (Fig.\ \ref{f:image}) and 
with a circle in the CMD (Fig.\ \ref{f:CMD}).

Taking into account that \HSTphot{} perfectly fits this object with stellar profile ($\textit{sharp}=-0.001$),
and assuming that the object belongs to the galaxy,
we can estimate an upper limit on linear size of \Halpha{} source.
Because the object profile is not wider than point spread function (PSF) we could assume that 
its angular size is smaller than the full width at half maxima (FWHM) for a stellar profile.
Thus, \Halpha{} source diameter should be less than 1.9 pc for WFPC2's $\textrm{FWHM}\sim0.2^{\prime\prime}$ 
at the distance 1.93~Mpc.
The median radius of old planetary nebulae (PN) is $\sim0.6$ pc, while young PN typically have radii $<0.05$ pc \citep{FP2010}.
Although, \hii{} regions have sizes from tens to hundreds parsecs \citep{O+2003}, 
the ultra compact \hii{} regions have diameters $<0.1$ pc \citep{WC1989}, that comparable in size with PN.

\begin{figure}
\centerline{
\includegraphics[width=0.45\textwidth]{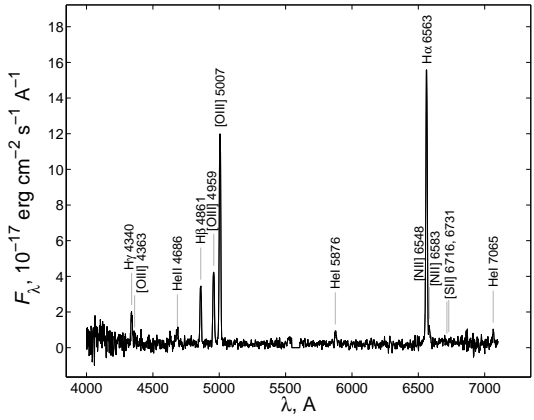}
}
\caption{The spectrum of the \Halpha{} object.}
\label{f:pnspec}
\end{figure}

Thus, intriguing question is the nature of the \Halpha{} emission in KKR\,25.
We carried out two sets of spectroscopic observations of the \Halpha{} object 
with the 6-meter telescope on 30th May 2009 and 8th August 2011 (see Table \ref{t:obslog}).
Each of 3 exposures of 1200 sec was obtained along the parallactic angle to minimize an influence of the atmospheric refraction.
The resulting spectrum is shown in Fig.\ \ref{f:pnspec}.
Besides bright hydrogen and oxygen emissions we clearly see helium (\HeI\,5876, 7065 and \HeII\,4686) and nitrogen ([$N\ii$]\,6548, 6583) lines.
Table~\ref{t:pnlines} presents measurements of the emission line intensities $F(\lambda)$ relative to H$\beta$,
and the ratios corrected for Galactic and internal extinction ($I(\lambda)/I(H\beta)$).

\begin{table}
\caption{Line intensities of the \Halpha{} object.}
\label{t:pnlines}
\scriptsize
\begin{tabular}{lccccr}
\hline
  Ion       & $\lambda$, \AA & $\frac{F(\lambda)}{F(H\beta)}$ &  $\frac{I(\lambda)}{I(H\beta)}$   \\ 
\hline                                                                          
 H$\gamma$  &   4340.47 & $0.52\pm0.08$ &  $0.66\pm0.13$ \\
 \OIII      &   4363.21 & $0.07\pm0.05$ &  $0.08\pm0.07$ \\
 He\ii      &   4685.70 & $0.30\pm0.07$ &  $0.32\pm0.08$ \\
 H$\beta$   &   4861.33 & $1.00\pm0.08$ &  $1.00\pm0.10$ \\
 \OIII      &   4958.92 & $1.36\pm0.10$ &  $1.30\pm0.10$ \\
 \OIII      &   5006.85 & $4.07\pm0.26$ &  $3.82\pm0.25$ \\
 \HeI       &   5875.60 & $0.23\pm0.05$ &  $0.16\pm0.04$ \\
 $[$N\ii$]$ &   6548.10 & $0.08\pm0.04$ &  $0.04\pm0.02$ \\
 \Halpha{}  &   6562.80 & $5.05\pm0.30$ &  $2.80\pm0.19$ \\
 $[$N\ii$]$ &   6583.40 & $0.26\pm0.06$ &  $0.14\pm0.03$ \\
 \HeI       &   7065.30 & $0.25\pm0.05$ &  $0.12\pm0.03$ \\
\hline                                                                                    
\end{tabular}
\end{table}

Because the object under consideration is a point-like source, we can not use morphological properties for classification.
Moreover, PN and \hii{} spectra have the same emission lines. 
The difference is only in intensity ratios due to different temperature of central source.
A central star in a PN is hotter than OB stars in \hii{} regions.
\citet{KPZ2008} proposed to use the following characteristic diagrams for the separation of PN and \hii{} regions 
(see formulas 1 and 3 from original publication):
\begin{equation}
\log \frac{\textup{\OIII}\,5007}{\textup{H}\beta} \ge (0.61/\log \frac{[\textup{N}\ii]\,6584}{\textup{H}\alpha} - 0.47) + 1.19
\label{e:c1}
\end{equation}
\begin{equation}
\log \frac{\textup{\SII\,6731,\,6717}}{\textup{H}\alpha} \le 0.63 \log \frac{\textup{[NII]\,6584}}{\textup{H}\alpha} - 0.55
\label{e:c2}
\end{equation}
These criteria allow to segregate 99 per cent of PNs from compact \hii{} regions if an object satisfies to at least one of these inequalities.
According to line intensity measurements (see Table~\ref{t:pnlines}) the \Halpha{} object passes both criteria.
The criterion~(\ref{e:c1}) is satisfied at a level of 12 sigma.
Because sulphur lines \SII\,6731, 6717 were not detected, we used $3\sigma$ upper limit for classification.
$\textrm{I}(\SII)/\textrm{I}(\textrm{H}\beta)<0.06$ meets the criterion~(\ref{e:c2}) at least on 4 sigma level.
Thus, we can argue with high reliability that \Halpha{} object is a planetary nebulae.

Using \Halpha{} flux $\log F = -14.64$ erg\,cm$^{-2}$\,s$^{-1}$ from \citet{CVnI+Halpha} 
and $F(\OIII)/F(\textrm{H}\alpha)$ ratio from our measurements (Table~\ref{t:pnlines}),
we can estimate visible $m_{5007}=-2.5\log F -13.74 = 23.09$ \citep{PNLF}, 
which corresponds to $M_{5007}=-3.35$ for a distance $(m-M)_0=26.42$ and $E(B-V)=0.008$.
Note, that the estimated absolute magnitude is close to bright cut-off of PN 
luminosity function. 

We serendipitously discovered a planetary nebula in KKR\,25.
PNs are not rare in giant galaxies. 
PN luminosity function is one of the popular methods for distance determination in the nearby Universe.
There are number of irregular galaxies with known PNs
as well as bright dwarf ellipticals NGC\,185, NGC\,205 near M\,31.
Up-to-date there are only 4 dwarf spheroidal system owning PN.
It is NGC\,147 \citep{G+2007}, a satellite of M\,31, and Fornax \citep{K+2007}, Sagittarius \citep{K+2008} 
dwarf spheroidals and transitional type Phoenix \citep{Saviane+2009}, are satellites of Milky Way.
Most of these galaxies are significantly brighter than KKR\,25 and contain substantially bigger stellar mass.
Only Sagittarius dSph and Phoenix are more less comparable with KKR\,25 by morphology and luminosity.
We have found first PN in dSph galaxy outside the Local Group.
An analysis of CMD on existence of extraordinary blue stars gives fast and easy method for selection of PN candidates in nearby galaxies.

We have estimated an oxygen abundance of PN to be $12+\log(\textrm{O/H}) = 7.60 \pm 0.07$, 
using the semi-empirical method of \citet{semiempirical}.
The line [\textrm{O}\,\textsc{ii}] 3727 is invisible on our data.
Thus, we estimated it as 2 sigma of noise in corresponding part of the spectrum.
This contributions to total abundance of oxygen is negligible and 
variations of its intensity in range $\pm1$ sigma do not change the result.

The PN also allows us to measure velocity of the galaxy with high precision,
which is really challenging for low surface brightness dwarf spheroidal galaxy without gas.
We have estimated the heliocentric velocity of the \Halpha{} object $V_h=-79 \pm 9$ \kms{}
using weighted mean of redshifts of emission lines (\Halpha{}, H$\beta$ and \OIII\,4959 and 5007) 
in each of 6 spectra obtained in 2009 and 2011 years with Russian 6-m telescope.

We have obtained spectra of \Halpha{} spot in KKR\,25 with the DEMOS spectrograph on the Keck\,II 10-m telescope. 
The observations were carried out during 2 nights on February 22 and 23, 2009.
We received $2\times5$ min spectra in the first and $3\times5$ min spectra in the second nights.
The spectroscopic setup for the DEIMOS observations used the 1200 lines\,mm$^{-1}$ grating  
with a central wavelength of 5800\AA. 
This provides a spectral coverage over a range of 4400 -- 7100 \AA{}, 
with the dispersion about 0.33\AA{} per pixel.
The observations took place against a very bright dawn sky.
Nevertheless, each 5 min spectral exposure shows a bright emission line close to \Halpha{} position.
A weak \textrm{N}\,\textsc{ii} 6548 line is also visible. There is only hint on \textrm{N}\,\textsc{ii} 6548.
We have estimated the heliocentric velocity $V_h=-54 \pm 12$ \kms{} using combined spectrum.

\section{Discussion}

\begin{table*}
\caption{
Properties of nearby isolated galaxies.
}
\label{t:gals}
\begin{tabular}{lrrrlrlllr}
\hline
Galaxy$^1$  &Type$^2$& MW M\,31 LG$^3$ & TI$^4$ & MD$^5$ &
\multicolumn{1}{c}{$M_V$$^6$}       & 
\multicolumn{1}{c}{$\Sigma_V$$^7$}  & 
\multicolumn{1}{c}{\feh$^8$}        & 
\multicolumn{1}{c}{$M$ $^9$}        & 
\multicolumn{1}{c}{$M/L_V$$^{10}$}  \\
            &   &\multicolumn{1}{c}{Mpc}&       &        &
\multicolumn{1}{c}{mag}             &
\multicolumn{1}{c}{mag}             &                
                                    & 
\multicolumn{1}{c}{$10^8 M_\odot$}  & 
\multicolumn{1}{c}{$M_\odot/L_\odot$} \\
\hline                                                                                                                                              
KKR\,25     & dSph& 1.93$^a$ 1.87 1.86 & $-1.0$ & M\,31  & $-10.93^a$           &$\phantom{\le\;}23.94^a$ &                &                          &                  \\[4pt]
Phoenix     & Tr  & 0.42$^b$ 0.86 0.58 & $0.7$  & MW     & $-9.95^c$            &$\phantom{\le\;}24.3^l$  &$-1.87\pm0.06^b$& $0.31^c$                 & 37               \\
Cetus       & dSph& 0.78$^d$ 0.69 0.62 & $0.3$  & M\,31  & $-11.34^e$           &$\phantom{\le\;}25.0^e$  & $-1.9^f$       & $1.1\phantom{0}\pm0.1^f$ & 37               \\
Tucana      & dSph& 0.88$^d$ 1.34 1.09 & $-0.2$ & MW     & $-9.78^g$            &$\phantom{\le\;}24.95^g$ & $-1.7\pm0.2^f$ & $0.43\pm0.15^f$          & 62               \\
And\,XVIII  & dSph& 1.36$^h$ 0.61 0.94 & $0.4$  & M\,31  & $-9.7^h\phantom{0}$  & $\le25.6^h$             & $-1.8\pm0.1^f$ & $0.27\pm0.15^m$          & 41               \\[4pt]
KKH\,98     & Irr & 2.49$^i$ 1.74 2.08 & $-0.9$ & M\,31  & $-12.23^j$           &$\phantom{\le\;}22.6^j$  & $-1.94^j$      & $0.66^a$                 & 10               \\
\hline
\multicolumn{10}{p{0.8\textwidth}}{
\textsc{Referencves} --
$^a$ this work;
$^b$ \citet{Hidalgo+2009};
$^c$ \citet{Mateo1998};
$^d$ \citet{CNG};
$^e$ \citet{MI2006};
$^f$ \citet{Kalirai+2010};
$^g$ \citet{SHP1996};
$^h$ \citet{McConnachie+2008};
$^i$ \citet{Melbourne+2010}
$^j$ \citet{Sharina+2008};
$^k$ \citet{FIGGS};
$^l$ $\Sigma_B=24.9$ \citep{ESOphot} has been transformed to $V$-band using $(B-V)_e=0.61$ \citep{PH1998};
$^m$ $M_{1/2}=2.7\times10^7$ $M_\odot$ \citep{Tollerud+2011} is the mass at the half-light radius of the And\,XVIII.
}\\
\multicolumn{10}{p{0.8\textwidth}}{
$^1$ name of the galaxy;\newline
$^2$ morphological type;\newline
$^3$ distance from the Milky Way (MW), the Andromeda galaxy (M\,31) and the centroid of the Local Group (LG) to the galaxy in Mpc;\newline
$^4$ tidal index;\newline
$^5$ main disturber;\newline
$^6$ $V$-band absolute magnitude of the galaxy;\newline
$^7$ $V$-band central surface brightness of the galaxy corrected for Galaxy absorption;\newline
$^8$ metallicity \feh of the galaxy;\newline
$^9$ total mass of the galaxy;\newline
$^{10}$ $V$-band total mass-to-luminosity ratio.
}
\end{tabular}
\end{table*}

We gathered the properties of KKR\,25, the Local Group spheroidals outside the virialized zones of Milky Way and M\,31,
and highly isolated dIrr KKH\,98 in Table~\ref{t:gals}. 
The columns contain the galaxy identification; its morphological type; distance in Mpc from the Milky Way (MW), 
the Andromeda galaxy (M\,31) and centroid of the Local Group (LG); tidal index (TI) and the main disturber as it defined
by \citet{CNG}; absolute magnitude $M_V$; central surface brightness $\Sigma_V$ corrected for Galactic absorption;
metallicity \feh; total mass $M$; and mass-to-light ratio $M/L_V$.
The original values were corrected for adopted distance if necessary.
The tidal index characterizes in logarithmic scale the tidal influence from the most important nearby galaxy.
The negative value corresponds to isolated object.
KKR\,25 is in $2.5$ -- $3$ times brighter than Phoenix, Tucana and And\,XVIII, and in 1.5 times fainter than Cetus -- the brightest isolated spheroidal.
KKR\,25 has the highest surface brightness of all spheroidal galaxies in our sample.

KKR\,25 is very similar to remote dwarf spheroidal galaxies in the Local Group.
Cetus, Tucana and And\,XVIII are the only dSphs which are not definitely satellites of the Milky Way or the Andromeda galaxy.
All these galaxies are gas deficient.
Any previous \hi{} detections were disproved by direct observations of stellar kinematics 
of Cetus \citep{LIC2007} and Tucana \citep{FTI2009},
or by deep radio interferometric observations of KKR\,25 \citep{KKR25+GMRT}.
Only in Phoenix, \citet{Gallart+2001} concluded that a \hi{} cloud was associated with the galaxy, 
but it has been lost after the last star formation episode. 
KKR\,25 resides near the brightest end of luminosity function of dwarf spheroidals.
It follows the same relation of surface brightness versus luminosity as satellites of Milky Way and M\,31.
We can expect that KKR\,25 traces the same properties as isolated dSphs in the Local Group and
has the similar mass and mass-to-light ratio.

In the Local Volume on the scale of a few Mpc we know only one galaxy comparable with KKR\,25 in isolation.
This highly isolated dwarf KKH\,98 is a normal dIrr galaxy.
\Halpha{} emission indicates ongoing star formation rate $\log(\textrm{SFR})=-3.5$ $M_\odot$\,yr$^{-1}$ (Kaisin, private communication).
The observations with GMRT in the framework of FIGGS project \citep{FIGGS} detect
$M_{HI}=6.46\times10^6$ $M_\odot$ of neutral hydrogen which spreads over 1.4 kpc from the centre of the galaxy. 
A \hi{} exceeds the optical size of KKH\,98 in 3.45 times.
A hydrogen mass-to-luminosity ratio $M_{HI}/L_V=1$ is typical for dwarf galaxies of given luminosity.
Using the data of \citet{FIGGS} ($W_{50}=20.7$ \kms{}, $i=46\degr$, $D_{\hi}=3.8^{\prime}$),
we estimated total indicative mass of KKH\,98 $M=6.6\times10^7$ $M_\odot$.
The obtained mass is similar to the mass of isolated dwarf spheroidals.
But KKH\,98 shows very different morphology, gas and star content.
It seems mysterious why the galaxies of similar mass and resemble environment have so different star formation history.

During the search of field dwarf galaxies without signs of recent star formation
in the luminosity range of $12>M_r>18$ mag and the stellar mass range of $10^7 <M_{\rm stellar}< 10^9$ \msol{}
\citet{Geha+2012} did not find such objects at distances above 1.5 Mpc from the central giant galaxy. 
Although KKR\,25 lies below the lower limits both in the stellar mass and the luminosity, but quite close to them.
KKR\,25 is one of the most isolated galaxies in the vicinity of the Local Group.
It does not contain a detectable amount of gas and can be reliably classified as dwarf spheroidal galaxy.
Obviously, models of galaxy formation should take into account an existence of such objects.
KKR\,25 stays far away from any massive galaxy in the Local Volume to be affected by an interaction during its evolution.
We can conclude that an evolution of KKR\,25 was regulated by star formation in the galaxy itself rather than by its environment.
The `primordial scenario' proposed that dwarf spheroidals form before the reionization in small haloes $M<2\times10^8$ $M_\odot$.
A star formation in these haloes is regulated by cooling and feedback processes in the early Universe.
Simulations of pre-reionization fossils explain main properties of dwarf spheroidals in the Local Group \citep{BR2009}.
It seems that KKR\,25 is the best candidate of such a `fossil' galaxy \citep{RG2005}.

\citet{TK2009} pointed out that the standard cosmological $\Lambda$CDM model predicts a factor of 10 more dwarf haloes in the field
than the number of observed dwarf galaxies. 
Thus the theory meet the same problem as overabundance of dwarf dark matter haloes in the Local Group,
so-called `missed satellites' problem.
\citet{TK2009} suggested several solutions.
One of it implies significant incompleteness of the observational sample of galaxies.
The model predicts 10 times more dwarf galaxies down to limiting magnitude $M_B=-12$ than listed in the \citet{CNG} sample.
\citet{TK2009} proposed that the dwarf spheroidal galaxies are good candidates for this role.
Indeed, the sample of dIrrs seems to be complete down to the limit of $M_B=-12$ \citep{CNG}.
They hold significant amount of \hi{} gas.
An ongoing star formation increases surface brightness and total luminosity of a galaxy.
Therefore, dIrrs are relatively easy to detect.
At the same time, dSphs are invisible in blind \hi{} surveys, 
they have very low surface brightness,
which makes them extremely difficult to find.
KKR\,25 satisfies all these conditions as a first member of population of `missed' galaxies.
It is an isolated, gas deficient, low surface brightness, dwarf spheroidal galaxy.
Taking into account only isolated spheroidal galaxies in the Local Group ($N\sim2$ -- $4$), 
we can roughly estimate a total number of similar systems inside the Local Volume up to 8 Mpc.
We could expect to find $N_t\sim8^3\times(2\textrm{--}4)\sim1000\textrm{--}2000$ objects
within the luminosity range $-9.5>M_V>-11.5$.
This number resembles the 1000 `missed' dwarf spheroidals mentioned by \citet{TK2009}.
But it is necessary to point out that KKR\,25 is one of the most luminous of known isolated dSphs.
It has $M_B\approx-10$ and lies well above the photometric limit of $M_B<-12$.
We could expect that other dSph galaxies will be even fainter.
Therefore, it seems unlikely that galaxies similar to KKR\,25 could solve the problem of `missed' dwarfs.
Anyway, the hunting for dwarf spheroidals in the field is an important task.
An existence of dSph population in voids is the crucial test on models of formation and evolution of dwarf galaxies.

\section{Conclusions}

We present a photometric and spectroscopic study of the unique isolated nearby dSph galaxy KKR\,25.
Let us briefly summarize the results of our study. 
We have estimated the distance modulus of KKR\,25 $(m-M)_0=26.42\pm0.07$ mag using the TRGB method.
It corresponds to a distance $D=1.93\pm0.07$ Mpc.
The new value is in good agreement with all previous distance measurements.

We have derived a quantitative star formation history of the isolated dwarf spheroidal galaxy KKR\,25.
The star formation history was reconstructed using \textit{HST}/WFPC2 images of the galaxy and 
a resolved stellar population modelling. 
According to our measurements, 62 per\,cent of the total stellar mass were formed during the initial burst of
star formation occurred about 12.6 -- 13.7 Gyr ago. 
There are indications of intermediate age star formation in KKR\,25 between 1 and 4 Gyr
with no significant signs of metal enrichment for these stars.

A distribution of the stars in the galaxy is well described by an exponential profile with central depression.
The exponential scale length is $h=156^{+12}_{-11}$ pc.
The profile extends up to 5 scale lengths.
The size of the depression $R=170^{+22}_{-30}$ is about the exponential scale length.

We did not confirm the presence of globular clusters in KKR\,25.
In the fact, the previously selected candidates are background objects, S0 galaxy at $z=0.34$ and QSO at $z=0.75$.

The spectroscopy of \Halpha{} object in KKR\,25 revealed that it is a planetary nebula 
with oxygen abundance $12+\log(\textrm{O/H}) = 7.60 \pm 0.07$.
We have serendipitously found the first PN in the dwarf spheroidal galaxy outside the Local Group.
The search of extraordinary blue stars on CMD of stellar population gives the 
perspective method for selection of PN candidates in distant galaxies.

We have derived heliocentric velocity of KKR\,25 using PN emission lines $V_h=-79\pm9$
and using integrated light of the galaxy $V_h=-65\pm15$.

Our study shows, that KKR 25 belongs to the population of highly isolated dwarf spheroidal galaxies, 
now rarely detected, but extremely important for our understanding of galaxy evolution theory.

The `primordial scenario' of galaxy formation is preferable against tidal stripping mechanism 
to explain the isolation of KKR\,25 and its morphology.
Existence of big number of dwarf spheroidals in the field could explain the 
overabundance problem in modern simulations.
The search for the dwarf spheroidal in voids is a crucial test for models of formation and evolution of dwarf galaxies.

\section*{Acknowledgements}
We are thankful to Dr.\ S.\ Pustilnik for very useful discussions of our work.
L.M.\ and D.M.\ are thankful to Else Starkenburg and  Tjitske Starkenburg for very interesting and valuable discussion.
We acknowledge the usage of the HyperLEDA database (\url{http://leda.univ-lyon1.fr}).
\textsc{stsdas} is a product of the Space Telescope Science Institute, which is operated by AURA for NASA.
The work was supported by the Russian Foundation for Basic Research (RFBR) grant 11--02--00639, 
Russian-Ukrainian RFBR grant 11--02--90449, 
and the program no. 17 `Active processes in galactic and extragalactic objects' of the Department of Physical Sciences of the Russian Academy of Sciences.
We acknowledge the support of the Ministry of Education and Science of the Russian Federation, 
the contract 14.740.11.0901.

\bibliographystyle{mn2e}
\bibliography{kkr25}   

\bsp
\label{lastpage}

\end{document}